\documentclass[pra,showpacs,twocolumn]{revtex4-1}

\usepackage[dvips]{graphicx}

\usepackage{amsmath}

\begin{document}

\title{Ground-states of spin-1 bosons in asymmetric double-wells}

\author{D. W. S. Carvalho, A. Foerster, M. A. Gusm\~ao}
\affiliation{Instituto de F\'isica, Universidade Federal do Rio Grande
  do Sul, C.P. 15051, 91501-970 Porto Alegre, Brazil}

\begin{abstract}

  In this work we investigate the different states of a system of
  spin-1 bosons in two potential wells connected by tunneling, with
  spin-dependent interaction. The model utilizes the well-known
  Bose-Hubbard Hamiltonian, adding a local interaction term that
  depends on the modulus of the total spin in a well, favoring a high-
  or low-spin state for different signs of the coupling constant. We
  employ the concept of \emph{fidelity} to detect critical values of
  parameters for which the ground state undergoes significant
  changes. The nature of the states is investigated through evaluation
  of average occupation numbers in the wells and of spin
  correlations. A more detailed analysis is done for a two-particle
  system, but a discussion of the three-particle case and some results
  for larger numbers are also presented.

\end{abstract}


\maketitle

\section{INTRODUCTION} \label{sec:intro} 

The experimental realization of Bose-Einstein condensates (BEC) in
dilute atomic gases \cite{Andetal95,Bradetal95,Davisetal95} is one of
the most exciting recent achievements in physics, research associated
to this peculiar state of matter has flourished in late years. One
remarkable development in this context is the realization of spinor
Bose gases in optical lattices.  In contrast to magnetic trap, where
spins are frozen, in optical trap the atoms keep their spin degrees of
freedom.  Several experimental groups have successfully created spinor
BECs of {$^{23}$Na} \cite{Stampetal98, Miesnetal99} and {$^{87}$Rb}
atoms \cite{Matthetal99, Barretal01, Changetal05}. Spinor gases
exhibit richer quantum effects than their single-component
counterparts, and allow to investigate mesoscopic magnetism.

These experimental developments have stimulated extensive study of
related theoretical models \cite{Ho98,Meleetal13, Meleetal12,
  Juliaetal09, Kuhnetal12a, Kuhnetal12b, Imambetal03, Tsuchetal04,
  Krutetal04, Kimuraetal05, Krutetal05, Paietal08, OhmiMachida98,
  Mustecetal07, Wagneretal11, Wagneretal12}. In particular, the
behavior of spin-1 bosons in a double-well potential can be described
by a variant of the two-site Bose-Hubbard Hamiltonian
\cite{jakschetal98} including spin-dependent interactions
\cite{Imambetal03, Wagneretal11} that affect physical properties of
the system.  Of particular interest is the case where the number of
atoms is small, motivated by the recent successful experimental
trapping of few atoms with high control and precision
\cite{Serwanetal11, Zurnetal12, Wenzetal13}.  Remarkably, the
experimental preparation of only two interacting particles in a double
well has been just reported \cite{Murmetal14}, and in principle the
extension to three particles is feasible by the state-of-the art
experiments \cite{Jochpriv}.  These experimental achievements have
generated an intense theoretical effort in few-body quantum systems
(see, for instance, \cite{Blume12, BraatHam06, Zollnetal08,
  Chattetal10, Brouz12, Harshman12, Harshman14, Wilsetal14,
  BrouFoers14, Zinnetal14, DAmicoRon14}). Despite their simplicity,
they still constitute a very challenging research field.

In this work we investigate the ground-state properties of few spin-1
bosons in a double well. Such systems can be viewed as building blocks
of optical lattices with cold atoms that can be in three different
hyperfine states. Besides the usual Hubbard-type repulsion, the model
includes a spin-dependent attractive interaction between the
particles, which may favor the establishment of a high or low spin
state in each of the wells depending on the sign of the coupling
constant of this interaction. Our basic goal is to study
changes in the characteristics of the ground state induced by
variations of the model parameters.

The paper is organized as follows. In Sec.~\ref{sec:model}, we discuss
the Hamiltonian and its diagonalization. A detailed analysis of
the two-particle system is developed in Sec.~\ref{sec:2part}, where we
also introduce the tools that we use to obtain information about
relevant properties. Sec.~\ref{sec:3pmore} shows relatively detailed
results for the three-boson system, and discusses general trends for
larger numbers, exemplifying with the cases of four and five
particles. Final remarks are presented in Sec.~\ref{sec:concl}.

\section{Model and matrix representation} \label{sec:model}

Following Ref.~\cite{Wagneretal11}, we write a variant of the
Bose-Hubbard Hamiltonian for a system composed of two wells, at
positions $L$ (left) and $R$ (right), as
\begin{align} \label{eq:hamil} 
  H &= \epsilon(n_L^{}-n_R^{}) - t \sum_{\sigma}(a_{L\sigma}^{\dagger}
  a_{R\sigma}^{} +   a_{R\sigma}^{\dagger} a_{L\sigma}^{}) \nonumber \\
  & \mbox{} + \tfrac{1}{2} \,U \sum_{i=L,R} n_i(n_i-1) + \tfrac{1}{2}
  \,U' \sum_{i=L,R}({\bf S}^2_i-2n_i),
\end{align} 
where $a_{i\sigma}^{\dagger}$ and $a_{i\sigma}^{}$ are the creation
and annihilation operators of a boson in a given well ($i = L, R$) and
in the spin state $\sigma = \{- 1,0,1 \}$; $n_i$ and ${\bf S}_i$ are
the number and total-spin operators for each well. We assume a single
level per well in the zero-tunneling limit ($t=0$), with energies
$\pm\epsilon$.  This means that $\epsilon$ is an asymmetry (or
\emph{tilt}) parameter, since the wells are identical only for
$\epsilon=0$. All interactions in the model are local. Apart from the
usual (repulsive) Hubbard interaction $U$, the last term in
Eq.~(\ref{eq:hamil}) describes a spin-dependent interaction with
coupling constant $U'$. It should be noted that this term contributes
only when more than one particles are present in the same well, in
which case low- and high-spin states are favored for $U' > 0$ and $U'
< 0$, respectively. The number and spin operators appearing in the
Hamiltonian are given by
\begin{equation} \label{eq:nS}
n_i = \sum_\sigma a_{i\sigma}^{\dagger}a_{i\sigma}\,, 
 \qquad
{\bf S}_i = \sum_{\sigma\sigma'} a_{i\sigma}^{\dagger} {\bf T}_{\sigma
  \sigma'} a_{i\sigma'}^{} \;, 
\end{equation} 
where ${\bf T} = {T}_x \hat {{\bf x}} +  {T} _y \hat
{{\bf y}} +  {T} _z \hat {{\bf z}} $, in terms of the usual
spin-1 matrices.

From Eq.~(\ref{eq:hamil}) we can see that the total number of
particles ($N_t = N_L + N_R$ ), the total spin (${\bf S}_t = {\bf S}_L
+ {\bf S}_R$), as well as any component of the latter (which we will
choose to be $S_{t}^z$) are conserved quantities.  It is natural to
investigate the properties of a system with fixed number of particles.
Moreover, since the energies do not depend on the value of $S_{t}^z$,
we can restrict our analysis to the subspace with $S_{t}^z$ eigenvalue
equal to zero, as this subspace is always present for any number of
spin-1 particles.

A convenient basis is provided by a set of vectors of the form
$|\{N_L^{}, N_R^{}\}, \{S_L^{}, S_R^{}\}, S_t^{} \rangle$, labeled by
the number and spin values in each well, and the total spin
$S_t^{}$. Bosonic symmetry imposes that $S_i+N_i$ must be an even
integer \cite{Wagneretal11} ($i=L,R$). This is an interesting basis
because it explicitly separates subspaces of different values of
$S_t^{}$, which are not connected by the Hamiltonian. However, to deal
with the tunneling term it is better to use appropriate symmetric
combinations of vectors of type $|N_L^{(0)}, N_L^{(1)}, N_L^{(-1)};
N_R^{(0)}, N_R^{(1)}, N_R^{(-1)} \rangle$, specifying the number of
particles per spin state in each well. Since these two sets of vectors
are related by Clebsh-Gordon coefficients, it is straightforward to
generate any matrix elements.

All results that we will discuss here were obtained starting from
exact diagonalization of the Hamiltonian matrix for a given set of
model parameters and particle number. The on-site Coulomb repulsion
$U$ was kept fixed, and chosen to be the energy unit. Hence, in what
follows we set $U=1$, it being implicitly assumed that any quantity
with dimension of energy is expressed in units of $U$.

\section{Two-particle system} \label{sec:2part}

We begin by investigating the energy spectrum for the simplest case,
$N_t=2$, with emphasis on changes in the ground state when varying the
parameters of the Hamiltonian.

\begin{figure}
\begin{center}
\includegraphics[width=8cm,clip]{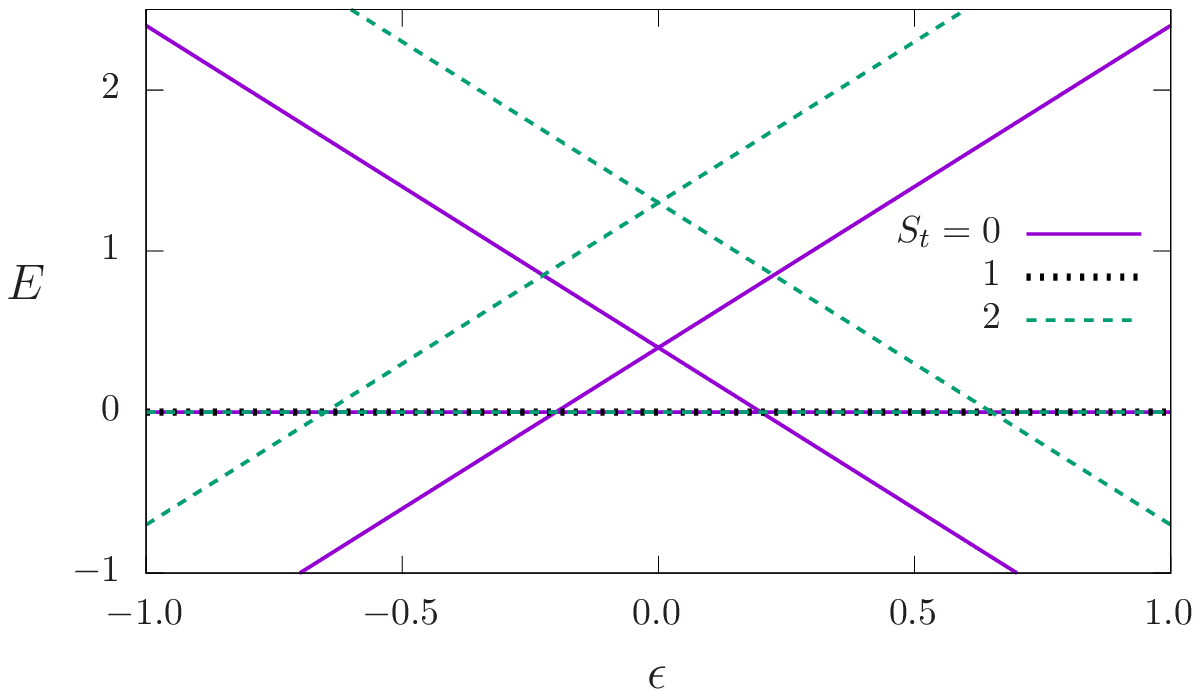}
\includegraphics[width=8cm,clip]{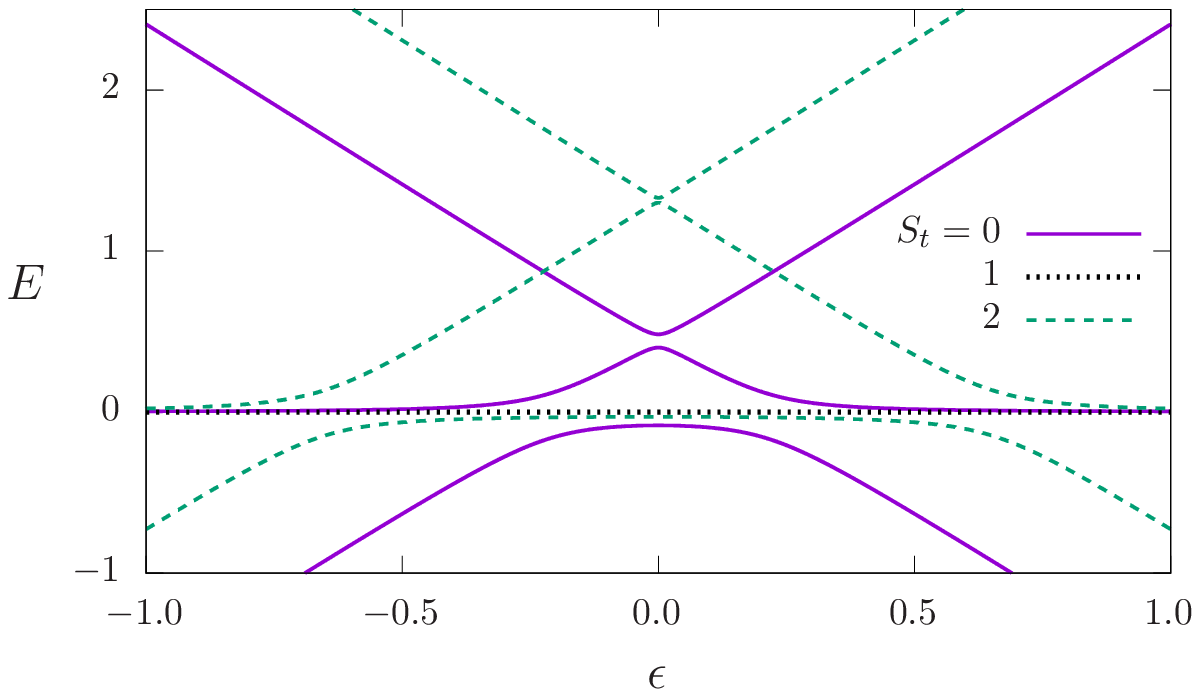}
\end{center}
\caption{(Color online) Energy eigenvalues as functions of $\epsilon$,
  for $N_t = 2$ and $U'=0.3$. The top panel shows decoupled wells
  ($t=0$), while the bottom panel corresponds to moderately weak
  tunneling ($t=0.1$).} \label{fig:energ}
\end{figure}

As an example, Fig.~\ref{fig:energ} shows the energy eigenvalues for
two bosons as functions of the asymmetry parameter $\epsilon$, for
moderately weak tunneling ($t = 0.1$) and spin-dependent coupling
($U'=0.3$). For comparison, the top panel shows the case of decoupled
wells ($t = 0$). We recall that the local energy levels are $\epsilon$
and $-\epsilon$ for left ($L$) and right ($R$) wells, respectively,
which can be related to their depths. Thus, $\epsilon < 0$ means that
the $L$ well is deeper. The inversion symmetry around $\epsilon = 0$
is noticeable in the plots.

From Fig.~\ref{fig:energ} it is clear that there are values of
$\epsilon$ near which the ground-state changes. These points occur at
level crossings in the absence of tunneling, and the degeneracy
lifting is stronger as tunneling increases.  

\subsection{Ground-state changes monitored by \emph{fidelity}}

Ground-state changes in finite-size systems may be viewed as precursors
of quantum phase transitions (QPTs) in the macroscopic limit.  Among
usual techniques to detect QPTs, the \emph{fidelity} of two
ground-states corresponding to different sets of parameters can be used
unambiguously for finite-size systems.  This is a concept derived from
quantum-information theory, and measures the similarity between two
quantum states. Although there are generalized definitions
\cite{NielsChuan04}, the simplest one, which serves our purposes,
is
\begin{equation} \label{eq:fid}
\mathcal{F}(\psi,\phi)=|\left<\psi|\phi\right>|\,.
\end{equation}
It defines the fidelity between any two states of the Hilbert space as
the absolute value of their scalar product. For normalized states, $0
\le \mathcal{F} \le 1$.

In our case, judging from Fig.~\ref{fig:energ}, an appropriate
control parameter is $\epsilon$, so that we define the fidelity
\begin{equation} \label{eq:fideps}
  \mathcal{F}_\epsilon(\epsilon;N_{t},t,U') =
  |\left\langle\epsilon-\delta,N_{t},t,U' |  \epsilon+\delta,N_{t},t,U'
  \right\rangle|\,,  
\end{equation}
where $\delta$ is small in the scale of $\epsilon$ values, and the
notation for the ground-state vector also includes all quantities that
are kept fixed. Later on we will be interested in the effect of
varying the spin-dependent interaction $U'$, in which case a more
convenient choice of fidelity is
\begin{equation} \label{eq:fidUp}
  \mathcal{F}_{U'}(\epsilon,N_{t},t;U') =
  |\left\langle\epsilon,N_{t},t,U'-\delta | 
    \epsilon,N_{t},t,U'+\delta \right\rangle|\,. 
\end{equation}

\begin{figure}
  \begin{center}
  \includegraphics[width=8cm,clip]{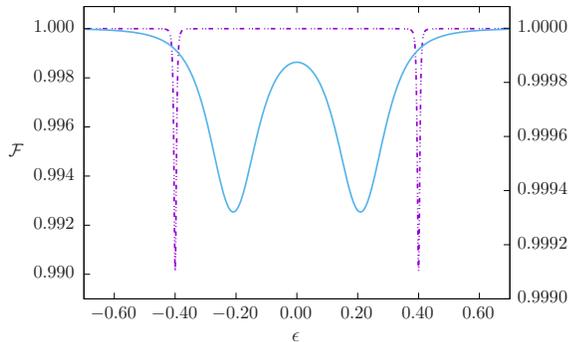} 
\end{center}
\caption{(Color online) Ground-state fidelity as a function of the
  asymmetry parameter $\epsilon$ for $N_t=2$, with two sets of values
  of tunneling amplitude and spin-dependent coupling: $t=0.1,\,U'=0.3$
  (solid line, right y-axis), and $t=0.005,\,U'=0.1$ (dashed, left
  y-axis).} \label{fig:fidN2}
\end{figure}

Characteristic behavior of $\mathcal{F_\epsilon}$ for two particles is
shown in Fig.~\ref{fig:fidN2}. The smoother curve corresponds to the
system parameters used in Fig.~\ref{fig:energ}. Clearly defined minima
are observed at values of the control parameter for which a
substantial change of ground-state occurs. These minima become much
sharper when the tunneling is reduced, as shown by the dashed curve,
for which they appear at different $\epsilon$ values because a
different $U'$ was chosen.

\subsection{Occupation number and spin correlations}

Although the fidelity finds parameter values for which the ground sate
changes, it does not give direct information about the nature of
states. For this we need to evaluate (average values of) relevant
physical quantities. An obvious one is the occupation number of each
well \cite{Wagneretal11}. Also interesting is the spin correlation
function, usually employed to study magnetic properties of solids, as
they are directly associated with magnetic susceptibility. The latter
not only gives the response to an applied field, but serves to signal
the establishment of magnetic order. The concept of magnetic order
does not make sense in a small system as the one that we are
studying. However, spin correlation functions can give important
information about the nature of the ground state with respect to
relative orientations of the spins.

We define the spin correlation function between the two wells as
\begin{equation} \label{eq:corr} C(\epsilon,N_t,t,U') = \left\langle
    \epsilon,N_t,t,U'|\mathbf{S}_L \cdot \mathbf{S}_R
    |\epsilon,N_t,t,U'\right\rangle,
\end{equation}
where $\mathbf{S}_L$ and $\mathbf{S}_R$ denote the total spin operator
associated to the left and right wells, respectively.

\begin{figure}
  \begin{center}
    \includegraphics[width=8cm]{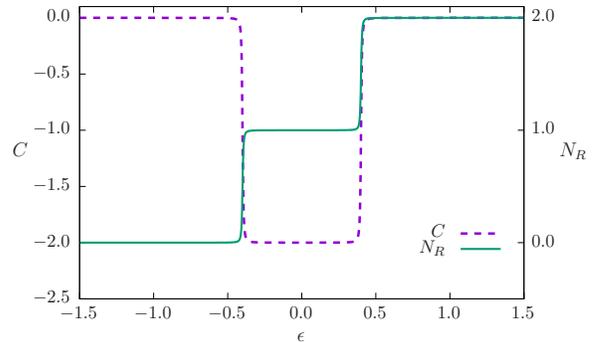}
\end{center}
\caption{(Color online) Occupation number of the $R$ well (solid line)
  and inter-well spin correlation function (dashed) as a function of
  $\epsilon$ for two particles, with the same parameters that show
  sharp minima in Fig.~\ref{fig:fidN2}.} \label{fig:NrCN2}
\end{figure}

Figure \ref{fig:NrCN2} shows typical behavior of the right-well
occupation number and inter-well spin correlation function for two
particles, in the same range of relative depths as in
Fig.~\ref{fig:fidN2}. The staircase behavior of $N_R$ is easy to
understand on the basis of a competition between the on-site
repulsion, that tends to keep particles apart, and the energy
asymmetry, that favors occupation of the deepest well. The role of $U'$
is to select spin states and to counteract the repulsion $U$ since it
is attractive for the appropriate spin orientations.

By the definition (\ref{eq:corr}), $C$ is zero if one of the wells is
empty or doubly occupied with zero total spin, which is observed in
the large $|\epsilon|$ regions of Fig.~\ref{fig:NrCN2}. In the
single-occupation regime, we see that the spin correlation is
negative, reflecting the fact that $U'>0$ favors a low total spin in
each well, which is consistent with tunneling to singly occupied
states with opposite spins.  This is similar to the mechanism of
exchange interaction between localized electrons of neighboring atoms
in a crystal, although in that case, due to the exclusion principle,
there is no need for a spin dependent interaction.

The results shown up to now correspond to a weak positive $U'$. From
Fig.~\ref{fig:fidN2} it is possible to see that the single-occupation
range shrinks as $U'$ grows.  This regime disappears when $U'$ exceeds
a critical value $U'_{+}=0.5$, since the attractive effect of $\;U'$
overcomes the repulsion $U$, favoring formation of a zero-spin pair
in the deeper well. 

In the case of $U' < 0$, still for a two-boson system, the spin
correlation behaves similarly to what was seen for $U'>0$ in
Fig.~\ref{fig:NrCN2}, except that the sign of $C$ is reversed. Double
occupancy continues to be favored by $U'$, but now with maximum total
spin in the doubly occupied well. Hence, the single-occupation region
for small $|U'|$ will show ferromagnetic correlations. The critical
value for suppression of this regime is now $U'_{-}=-1.0$ (meaning
 $U'_{-}=-U$).

 \subsection{Two-particle regimes in the $(\epsilon,U')$ 
   plane}

\begin{figure}
  \centering
 {\includegraphics[width=8cm]{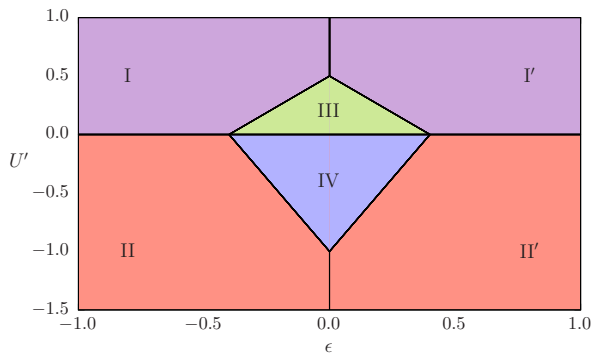}}
 \caption{(Color online) Regions of qualitatively different
   ground-states for two particles. The outer regions correspond to
   double occupation of the $L$ (I,II) and $R$
   ($\mathrm{I}'$,$\mathrm{II}'$) wells. Single occupation of both
   wells occur in the center region, with AF (III) and FM (IV) spin
   correlations.} \label{fig:mapN2}
\end{figure}

 It is interesting to see a complete picture of the case $N_{t} = 2$,
 varying the asymmetry parameter $\epsilon$ and the spin-dependent
 interaction $U'$, as shown in Fig.~\ref{fig:mapN2}.  The limiting
 lines of the various regions were obtained from fidelity minima
 (mostly $\mathcal{F}_\epsilon$, but the line at $U'=0$ is better seen
 with $\mathcal{F}_{U'}$). In regions I and II there is double
 occupation of the $L$ well, with $S_L = 0$ and 2,
 respectively. Regions $\mathrm{I}'$ and $\mathrm{II}'$ are equivalent
 to I and II, but the particles are located in the $R$ side. States of
 one particle in each well are observed in the two central regions,
 III and IV, respectively with negative and positive spin
 correlations. Borrowing denominations from magnetism, we can say that
 spin correlations are antiferromagnetic (AF) in region III and
 ferromagnetic (FM) in IV.

\begin{figure}
  \centering
 {\includegraphics[width=8cm]{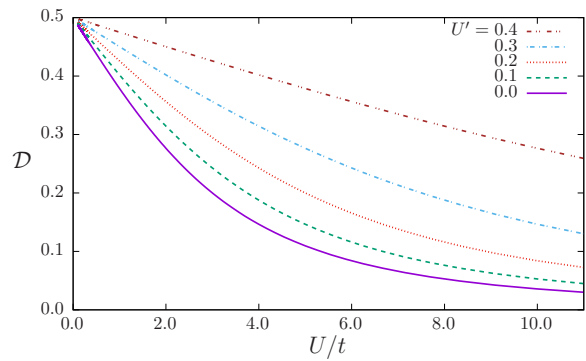}}
 \caption{(Color online) Variation of the average double-occupancy as
   a function of the the ratio between Hubbard repulsion and tunneling
   strength in the two-particle ground-state for a symmetric
   double-well. The curves correspond to values of the spin-dependent
   interaction within the region III of Fig.~\ref{fig:mapN2}.}
\label{fig:doubleocc}
\end{figure}

It is worth observing that the characterization of states in terms of
doubly or singly occupied wells refers to the average
occupation. Fluctuations occur for any $t \ne 0$, so that the sharp
boundary lines of Fig.~\ref{fig:mapN2} are only sharp for very weak
tunneling. For instance, choosing a favorable condition for single
occupancy inside region III (e.g, symmetric wells and small $|U'|$), a
possible measure of the probability of double occupancy is given by
$\mathcal{D} \equiv 1 - \mathcal{F}_t^2$, where $\mathcal{F}_t$ is the
fidelity between the ground-states with zero and nonzero
tunneling. This is shown as a function of $U/t$ in
Fig.~\ref{fig:doubleocc}, where we can see that it is null for
decoupled wells, and grows with the tunneling strength, approaching
$1/2\,$ for large $t$, when all possible occupancies become
energetically equivalent. The bottom curve ($U'=0$) is very similar to
the one obtained for two fermions in a double-well
\cite{Murmetal14}. Here, the effect of increasing $U'$ is to lower the
energy of doubly occupied wells.

\section{Three or more particles} \label{sec:3pmore}

\begin{figure}
 \begin{center}
\includegraphics[width=7.2cm,clip]{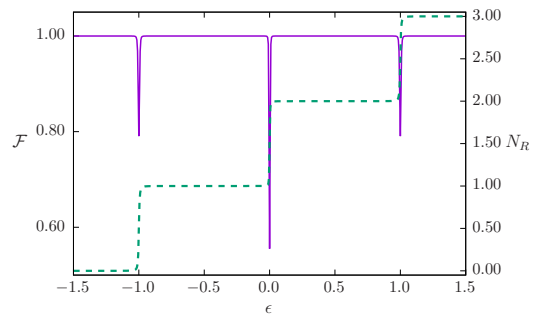} \\
\end{center}
\caption{(Color online) Fidelity (continuous) and $R$-well occupation
  (dashed) for $N_{t}=3$, $t=0.005$, and $U'=0.1$}\label{fig:fidNrN3}
\end{figure}

\begin{figure}
 \begin{center}
\includegraphics[width=7.2cm,clip]{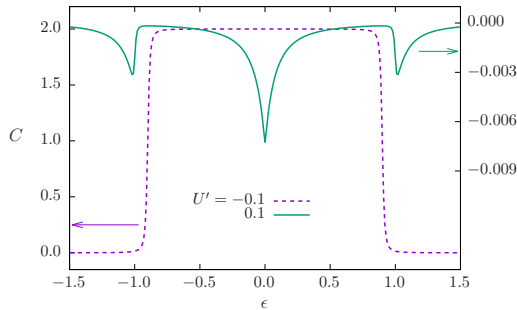}
\end{center}
\caption{(Color online) Spin correlation functions for $N_{t}=3$ and
  $t=0.005$ in the cases of positive and negative $U'$. Notice the
  enlarged scale and displaced origin for the solid
  line.} \label{fig:corrN3}
\end{figure}

Adding a third particle to the system leads to noticeable qualitative
changes. Figure~\ref{fig:fidNrN3} shows the fidelity and $R$-well
occupation for varying $\epsilon$. The plot is for $U'>0$, but the
behavior for negative $U'$ is qualitatively the same. Notice that
there is a central minimum of the fidelity, corresponding to the extra
step observed for $N_R$, but no central plateau, in contrast with
Fig.~\ref{fig:fidN2}, since it is impossible to have the same number
of particles in both wells. The most striking difference comes from
the spin correlation function, shown in Fig.~\ref{fig:corrN3}. For
positive $U'$ the correlation is negative but close to zero except,
for a slight increase in magnitude near the changes of
ground-state. This is due to the fact that there is always either zero
or two particles in one of the wells, which means that one of the
spins is always essentially null. In contrast, for $U'<0$ the whole
central region, where none of the wells is empty, presents a large
positive correlation between the spins of the two wells.

\begin{figure}
  \begin{center}
 {\includegraphics[width=8.5cm,clip]{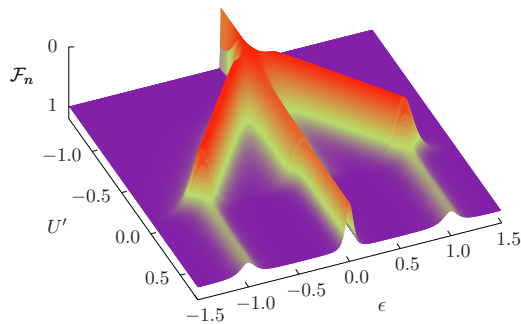}}
\end{center}
\caption{(Color online) Variation of the ground-state fidelity with
  $\epsilon$ and $U'$ for $t=0.1$ in a three-particle system. Notice
  that the fidelity axis is inverted, and the values are normalized to
  fall in the range $[0,1]$.} \label{fig:fid3dN3}
\end{figure}

As in the two-particle case, when the magnitude of a negative $U'$
grows beyond a critical value (once more, $U'_{-}=-1$), all three
particles stick together in one of the wells (with maximum spin
$S_i=3$), so that the two inner regions around the central minimum in
Fig.~\ref{fig:fidNrN3} cease to exist. On the other hand, for positive
$U'$ the fidelity minima are located at the same $\epsilon$ values
seen in Fig.~\ref{fig:fidNrN3}, independently of $U'$. This is shown
in Fig.~\ref{fig:fid3dN3}, where we plot a \emph{normalized} fidelity
$\mathcal{F}_n$ (rescaled to fall in the range $[0,1]$) as a function
of both $\epsilon$ and $U'$. The observed independence on $U'$ is due
to the constraint that $N_i+S_i$ must be even, which means that the
minimum total spin in a given well is $S_i=0$ for $N_i=2$ and $S_i=1$
for $N_i=3$. It then follows that the eigenvalue associated with the
$U'$ term [see Eq.~\eqref{eq:hamil}] is $S_i(S_i+1)-2N_i=-4$ in both
cases, so that the change in ground-state is driven by a balance
between $\epsilon$ and the local repulsion $U$.

\begin{figure}
  \centering
 {\includegraphics[width=8cm]{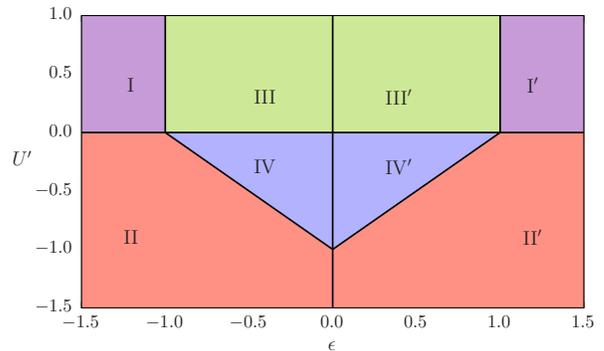}}
 \caption{(Color online) Regions of qualitatively different
   ground-states for three particles. The outer regions correspond to
   double occupation of the $L$ (I,II) and $R$
   ($\mathrm{I}'$,$\mathrm{II}'$) wells. In the central regions (III,
   $\mathrm{III}'$, IV, $\mathrm{IV}'$) there is single occupation in
   one well and double in the other, with very weak AF correlations
   for $U'>0$ and strong FM correlations for
   $U'<0$.} \label{fig:mapN3}
\end{figure}

A ground-state diagram like that of Fig.~\ref{fig:mapN2} can be
built for $N=3$, as shown in Fig.~\ref{fig:mapN3}. Here too the
regions labeled with primed roman numbers are equivalent to the
corresponding unprimed ones upon the exchange $L \leftrightarrow
R$. In I and II we have essentially three particles in $L$ while $R$
is nearly empty, with $S_t=3$ in I and $S_t=1$ in II. Regions III and
IV correspond to double occupation of $L$ and the third particle in
$R$, with the spin states compatible with spin correlations as shown
in Fig.~\ref{fig:corrN3}.

\subsection{Larger $N$}

\begin{figure*}
  \begin{center}
    \includegraphics[width=8.5cm]{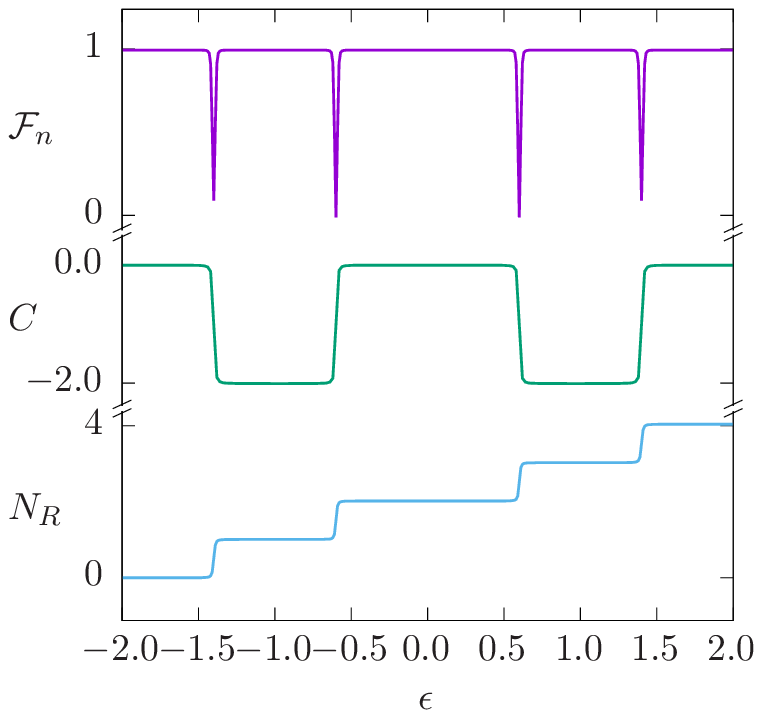} 
    \includegraphics[width=8.5cm]{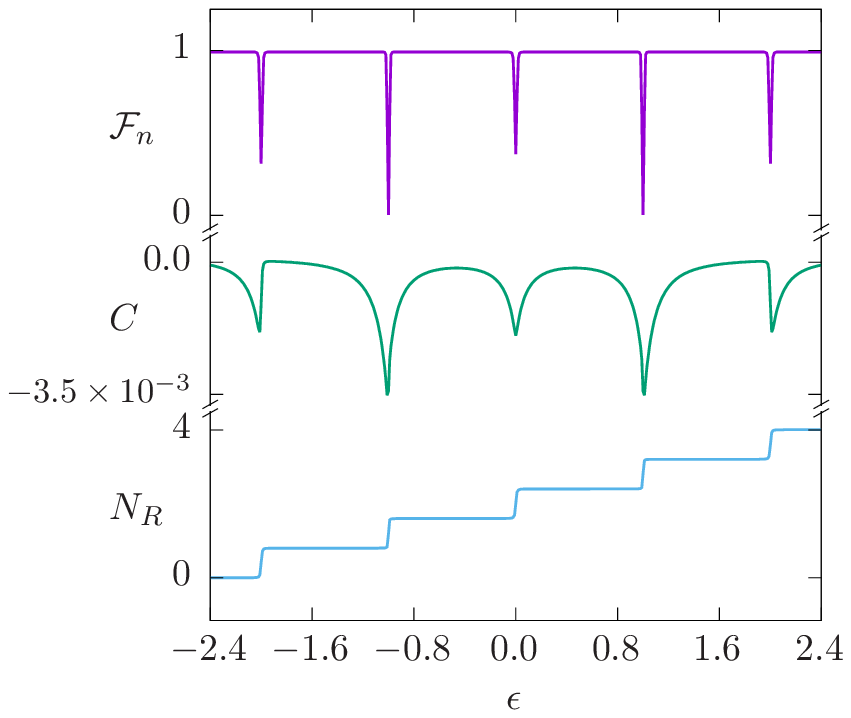}
  \end{center}
  \caption{(Color online) Normalized fidelity ($\mathcal{F}_n$), spin
    correlation function ($C$), and $R$-well occupation for systems of
    four particles (left panel) and five particles (right panel), with
    $t=0.005$ e $U'=0.1$.}
  \label{fig:fidNrC_N45}
\end{figure*}

The main trends in behavior observed for two and three particles
appear in general for even and odd particle number. This is
exemplified in Fig.~\ref{fig:fidNrC_N45} for $N_t=4$ and 5.
Obviously, the staircase increase in single-well average occupation
has a number of steps reflecting the total particle number. Each step
is accompanied by a sharp minimum of the fidelity.

Spin correlation functions are shown in Fig.~\ref{fig:fidNrC_N45} only
for $U'>0$. They reproduce the scenarios already viewed for two and
three particles. Also similarly to those cases, correlations for
$U'<0$ are positive (when not null), and have significant values, that
remain nearly constant between jumps of single-well occupations. 

The ``phase diagrams'' are similar to Fig.~\ref{fig:mapN2} for even
$N_t$ and to Fig.~\ref{fig:mapN3} for odd $N_t$, except that the
central part has a growing number of nested regions as the number of
possible distributed occupations increases.

\section{Conclusions} \label{sec:concl}

We studied the possible ground states of spin-1 bosons in double-well
potentials. Such systems can model the basic unit of optical lattices
with trapped cold atoms, for which parameters such as the depth of the
wells, amplitude of tunneling between them and interactions between
particles can be controlled. We based our analysis on the usual
Bose-Hubbard Hamiltonian with an additional on-site spin-dependent
interaction, as proposed in Ref.~\cite{Wagneretal11}.

Even considering different total numbers of particles, restriction to
two wells allowed us to exactly diagonalize the Hamiltonian matrix in
relevant subspaces, and to study changes of the ground state induced
by variation of the model parameters. Our analysis focused primarily
on a regime of weak tunneling relative to the local repulsive
(Hubbard) interaction, which we kept fixed. The variable parameters
were then the relative depth between the two wells and the local
spin-dependent coupling, which we allowed to be positive or negative,
respectively favoring low and high total spin in each well.

We showed that regime changes in the system can be detected by
evaluating the ground-state fidelity as some parameter is varied. This
quantity presents sharp minima at parameter values for which the
nature of the ground state changes, which should correspond to
critical values for occurrence of quantum phase transitions in the
macroscopic limit. Information on the nature of the different ground
states has to be sought through evaluation of average values of
appropriate physical quantities, like the number of particles and
total spin in a given well, or inter-well spin correlations.

The results allowed us to construct maps of different regimes in
parameter space, identifying regions of full occupancy of a single
well, and regions with particles distributed in both wells, in which
case either ferromagnetic or antiferromagnetic spin correlations
between the two wells occur as a consequence of the spin-dependent
interaction. This was done in detail for system of two and three
particles. Comparison with some results for four and five particles
revealed that the main qualitative differences occur between even and
odd total number of particles in the system.

This study can be extended to larger numbers, not only of particles
but also of wells, the limitations being only computational.  Work in
dynamical processes in the same model is now in progress, including
investigation of spin effects on transistor-like behavior
\cite{Sticknetal07} in the three-well case. Additionally, by
increasing the number of wells we can address the problem of Anderson
localization \cite{Cestarietal10} with a distribution of well depths,
taking into account the effect of nonzero spin and spin-dependent
interactions.

\acknowledgments

A. F. thanks A. Wagner for sharing important information, and
acknowledges invaluable discussions with S. Jochim.  This work was
supported in part by CAPES (Coordena\c c\~ao de Aperfei\c coamento de
Pessoal de N\'ivel Superior) and by CNPq (Conselho Nacional de
Desenvolvimento Cient\'ifico e Tecnol\'ogico), Brazil.

\end{document}